\documentclass[times,preprint]{aastex631}
\usepackage{CJK}
\usepackage{upgreek}
\usepackage{amsmath}
\usepackage{booktabs}

\newcommand{\mach}{\mathcal{M}}

\shorttitle{GW from PBH Inspiraling inside Compact Star}
\shortauthors{Zou \& Huang}

\begin{document}
\begin{CJK*}{UTF8}{gbsn}

\title{Gravitational-wave Emission from a Primordial Black Hole Inspiraling inside a Compact
       Star:\\ a Novel Probe for Dense Matter Equation of State}

\correspondingauthor{Yong-Feng Huang}
\email{hyf@nju.edu.cn}

\author[0000-0002-6189-8307]{Ze-Cheng Zou (邹泽城)}
\affiliation{School of Astronomy and Space Science, Nanjing University, Nanjing 210023, China}

\author[0000-0001-7199-2906]{Yong-Feng Huang (黄永锋)}
\affiliation{School of Astronomy and Space Science, Nanjing University, Nanjing 210023, China}
\affiliation{Key Laboratory of Modern Astronomy and Astrophysics (Nanjing University),
             Ministry of Education, Nanjing 210023, China}

\begin{abstract}

Primordial black holes of planetary masses captured by compact stars are widely studied to
constrain their composition fraction of dark matter. Such a capture may lead to an inspiral
process and be detected through gravitational-wave signals. In this Letter, we study the
postcapture inspiral process by considering two different kinds of compact stars, i.e.,
strange stars and neutron stars. The dynamical equations are numerically solved, and the
gravitational wave emission is calculated. It is found that the Advanced LIGO can
detect the inspiraling of a $10^{-5}$ solar mass primordial black hole at a distance of 10
kpc, while a Jovian-mass case can even be detected at megaparsecs. Promisingly, the next
generation gravitational wave detectors can detect the cases of $10^{-5}$ solar mass
primordial black holes up to $\sim1\,\mathrm{Mpc}$, and can detect Jovian-mass cases
at several hundred megaparsecs. Moreover, the kilohertz gravitational wave signal shows
significant differences for strange stars and neutron stars, potentially making it a novel
probe to the dense matter equation of state.

\end{abstract}


\section{Introduction} \label{sec:intro}
\end{CJK*}

Primordial black holes (PBHs; \citealt{1967SvA....10..602Z}) can
be formed through a vast range of mechanisms. For example, they
may be generated due to the density inhomogeneity in the early
universe \citep[for a review, see][]{2010RAA....10..495K}.
Although having not been directly detected yet, PBHs are
considered to be a candidate for dark matter \citep[for a review,
see][]{2020ARNPS..70..355C}. To constrain PBHs' composition
fraction of dark matter, the event rate of collisions between
planetary-mass PBHs and compact stars has been widely discussed
and the corresponding electromagnetic emissions have been
extensively studied
\citep[e.g.,][]{2013PhRvD..87l3524C,2014JCAP...06..026P,2017PhRvL.119f1101F,2018ApJ...868...17A,2020PhRvD.102h3004G}.
Interestingly, during such a collision, the PBH-compact star
system will also emit gravitational waves (GWs) as the PBH
accretes matter from the compact star. \citet{2016PhRvD..93b3508K}
and \citet{2019PhRvD.100l4026E} studied the GW signals from an
accreting PBH, which is assumed to have plunged into its compact
companion and stays at the exact center of the neutron star (NS).
\citet{2019PhRvL.122g1102H} studied the GW emission during
the inspiraling process of a PBH inside an NS, but their treatment
of the dynamics and the compact star structure is still
preliminary. \citet{2020PhRvD.102h3004G} analyzed the process of a
PBH plunging into an NS by considering dynamical friction,
accretion, and GW emission. For a trapped PBH, they found that
both the frequency and amplitude of the GWs are constant during
the inspiral. However, their dynamical equations are mainly
appropriate for PBHs in deeply subsonic motions. Also, they
assumed that the NS has a homogeneous structure when calculating
the motion of the trapped PBH, causing additional deviation in the
derived GW waveforms.

On the other hand, the equation of state (EoS) of dense matter determines the structure of
compact stars. According to the strange-quark matter
hypothesis \citep{1970PThPh..44..291I,1971PhRvD...4.1601B,1984PhRvD..30.2379F,1984PhRvD..30..272W},
pulsars may actually be strange stars (SSs) consisting of strange-quark matter \citep{1986ApJ...310..261A}.
The strange-quark matter is self-bound, thus strange dwarfs and even strange planets can stably exist.
However, a $1.4\,M_\odot$ SS has a radius very similar to that of a normal NS with
comparable mass, thus it is hard to distinguish between these two types of compact stars via
observations \citep{2015ApJ...804...21G,2021Innov...200152G}. An interesting method to identify
strange quark objects is to search for very close-in binary systems containing a strange planet and
a compact star \citep{2015ApJ...804...21G,2019AIPC.2127b0027K,2020ApJ...890...41K,2021arXiv210915161W}.
When the orbital radius of a planet is less than $\sim 5.6 \times 10^{10}$ cm or the orbital period
is less than $\sim 6100$ s, then it cannot be a normal matter planet, but should be a strange
planet, because the tidal force is too strong to allow any kind of normal matter planets to
stably exist there \citep{2017ApJ...848..115H,2019AIPC.2127b0027K,2020ApJ...890...41K}.
However, in these close-in ``planetary'' systems, there is still a possibility that the planetary-mass
object is actually a PBH. We thus need to further scrutinize its nature.

GW astronomy may shed new light on the study of compact star structure. Tidal deformability and maximal
mass of compact stars, which can be hinted at through GW signals from double compact star
mergers \citep[for a review, see][]{2019JPhG...46l3002G}, may reflect the internal composition and
structure of compact stars \citep[e.g.,][]{PhysRevD.104.123028}. In fact, the observations of GW170817
have already put useful constraints on the tidal deformability of
``neutron stars'' \citep{2018PhRvL.120q2703A,2018PhRvL.121p1101A,2019PhRvX...9a1001A}. However, these
constraints are still too weak to pin down the
EoS \citep{2019EPJA...55...60L,2019PhRvD.100b3015S,2021arXiv211104520M}.
In this Letter, we will study the GW emission produced by a PBH inspiraling inside a compact star.
Different from a strange planet, a PBH can inspiral and tunnel inside the compact star, producing
special GW signals. Two kinds of EoSs will be assumed for the compact stars in our calculations,
i.e., a strange-quark matter EoS and a hadronic matter EoS. The results will help us judge whether
the close-in planetary-mass object is a strange planet or a PBH.

The structure of this Letter is as follows. In
Section~\ref{sec:model}, we describe the model by setting up the
equations of motion and compact star structure. Numerical results
on the dynamics are then presented in Section~\ref{sec:binary}. In
Section~\ref{sec:gw}, the GW emissions are calculated and
analyzed. The effects of orbital eccentricity are
considered in Section~\ref{sec:e}, and the event rate is
calculated in Section~\ref{sec:rate}. Finally, we summarize and
discuss our results in Section~\ref{sec:conclude}. Throughout this
Letter, a natural unit system of $c=G=\hbar=1$ is adopted.

\section{Model Setup}\label{sec:model}

As a PBH inspirals inside a compact star, it gradually loses its
orbital kinetic energy through interactions with compact star
matter. We focus on three main channels --- dynamical friction,
accretion, and GW emission \citep{2020PhRvD.102h3004G}.
Moreover, we assume that the trajectory of the PBH within
the compact star is quasi-circular and do not consider the effects
of ellipticity in most of our calculations. This condition can be
guaranteed by the tidal dissipation \citep{2014ARA&A..52..171O}
and various circularization effects \citep{2013ApJ...774...48M}.
Anyway, a brief discussion on the effects of eccentricity will be
presented in Section~\ref{sec:e}.

\subsection{Dynamical Friction}

A wake is produced when the PBH (of mass $m_\mathrm{PBH}$) moves through the surrounding medium
(with a density of $\rho$ and the sound speed $c_\mathrm{s}$). The wake will then exert a gravitational
drag force on the PBH, known as the dynamical friction force \citep{1999ApJ...513..252O}. For a circular
orbit, the force can be written on the basis of the radial ($\hat{\boldsymbol{r}}$) and
azimuthal ($\hat{\boldsymbol{\varphi}}$) components as \citep{2007ApJ...665..432K}
\begin{equation}
  \boldsymbol{F}_\mathrm{DF}=-\frac{4\uppi\rho m_\mathrm{PBH}^2}{v^2}
  \left(\mathcal{I}_r\hat{\boldsymbol{r}}+\mathcal{I}_\varphi\hat{\boldsymbol{\varphi}}\right),\label{eq:Fdf}
\end{equation}
where $v$ is the relative speed of the PBH with respect to the medium. The coefficients $\mathcal{I}_r$
and $\mathcal{I}_\varphi$ are functions of the Mach number ($\mathcal{M}=v/c_\mathrm{s}$) and the distance ($r$)
between the PBH and the center of the compact star \citep{2007ApJ...665..432K}:
\begin{equation}
  \mathcal{I}_r=\begin{cases}
    \mach^210^{3.51\mach-4.22},&\mach<1.1;\\
    0.5\ln\left[9.33\mach^2\left(\mach^2-0.95\right)\right],&1.1\leqslant\mach<4.4;\\
    0.3\mach^2,&4.4\leqslant\mach;
  \end{cases}
\end{equation}
and
\begin{equation}
  \mathcal{I}_\varphi=\begin{cases}
    0.7706\ln\left(\frac{1+\mach}{1.0004-0.9185\mach}\right)-1.4703\mach,&\mach<1.0;\\
    \ln\left[330\left(r/r_\mathrm{m}\right)\left(\mach-0.71\right)^{5.72}\mach^{-9.58}\right],&1.0\leqslant\mach<4.4;\\
    \ln\left(\frac{r/r_\mathrm{m}}{0.11\mach+1.65}\right),&4.4\leqslant\mach;
  \end{cases}\label{eq:Iphi}
\end{equation}
where $r_\mathrm{m}=\sqrt{\mathrm{e}}m_\mathrm{PBH}/(2v^2)$ \citep{2011MNRAS.418.1238C}. However, the
above expression of $\mathcal{I}_\varphi$ fails for $\mach\to0$ because it gives an unphysically positive
azimuthal drag force. In fact, in this deep subsonic phase, the circular-orbit dynamical friction is expected
to be similar to the case of a straight trajectory \citep{1999ApJ...513..252O}, i.e.,
$\mathcal{I}_\varphi=\ln\left[(1+\mach)/(1-\mach)\right]/2-\mach\to\mach^3/3$. To account for
this asymptotic behavior, we use the following polynomial expansion for $\mach\ll1$,
\begin{equation}
  \mathcal{I}_\varphi=\mach^3/3-0.80352\mach^4+7.68585\mach^5. \label{eq:df_mach}
\end{equation}
This expression reduces to the solution of \citet{1999ApJ...513..252O} for $\mach\to0$. When $\mach$ increases,
it smoothly transitions to Equation~\ref{eq:Iphi} at $\mach=0.08588$. So when $\mach<0.08588$,
Equation~\ref{eq:df_mach} is applied instead of Equation~\ref{eq:Iphi} in our calculations.

\subsection{Accretion}

As the PBH accretes matter from the compact star, it also accumulates a negative momentum,
resulting in a drag force of \citep{2004NewAR..48..843E,2020PhRvD.102h3004G}
\begin{equation}
  \boldsymbol{F}_\mathrm{acc}=-\dot{m}_\mathrm{PBH}\boldsymbol{v}.\label{eq:Facc}
\end{equation}
For a PBH with a finite velocity, the accretion rate is \citep{1952MNRAS.112..195B,1985MNRAS.217..367S,2004NewAR..48..843E}
\begin{equation}
  \dot{m}_\mathrm{PBH}=\frac{4\uppi\lambda\rho m_\mathrm{PBH}^2}{(c_\mathrm{s}^2+v^2)^{3/2}},\label{eq:accrete}
\end{equation}
where $\lambda$ is the accretion eigenvalue depending on the EoS of the accreted matter.
Equation~\ref{eq:accrete} is applicable for $m_\mathrm{PBH}\ll M_\mathrm{CS}$ \citep{2021PhRvD.103j4009R},
where $M_\mathrm{CS}$ is the mass of the compact star.

\subsection{Structure of Compact Stars}\label{subsec:CS}

\begin{figure}
  \plotone{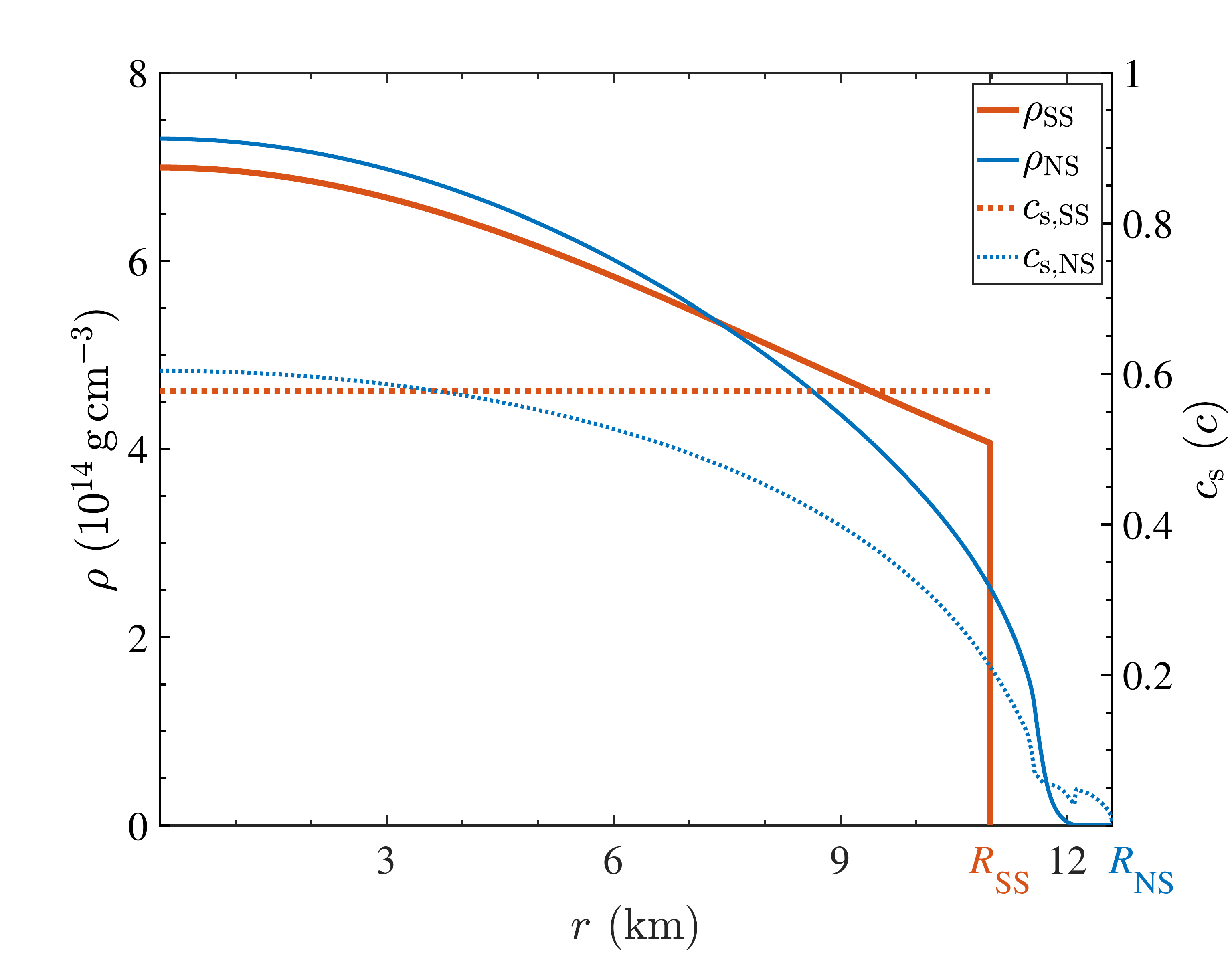}
  \caption{Density (solid curves; left $y$-axis) and sound speed (dotted curves; right $y$-axis)
           profiles of a $1.4\,M_\odot$ strange star (SS, thick curves) and neutron star
           (NS, thin curves).\label{fig:cs}}
\end{figure}

The dynamical friction and accretion explicitly depend on $\rho$ and $c_\mathrm{s}$. Therefore,
we shall obtain the compact star structure before further studying the equations of motion of the PBH.
The structure of a non-rotating compact star is available by solving the Tolman-Oppenheimer-Volkoff
equation \citep{1939PhRv...55..364T,1939PhRv...55..374O}
\begin{equation}
  \frac{\mathrm{d}P}{\mathrm{d}r}=-\frac{M_r}{r^2}\rho\left(1+\frac{P}{\rho}\right)
  \left(1+\frac{4\uppi r^3P}{M_r}\right)\left(1-\frac{2M_r}{r}\right)^{-1},\label{eq:TOV}
\end{equation}
where $P$ is the pressure, and $M_r$ is the mass within the radius $r$ so
that $\mathrm{d}M_r/\mathrm{d}r=4\uppi r^2\rho$. To solve Equation~\ref{eq:TOV}, one needs the EoS
of compact star matter. Two typical kinds of compact stars, SSs and normal NSs, are considered in our
study. For SSs, we adopt the simple bag model with massless quarks \citep{1984PhRvD..30.2379F}, whose
EoS reads $P=(\rho-4B)/3$. The bag constant $B$ is taken as 57 MeV\,fm$^{-3}$. For NSs, we use the hadronic
BSk 24 EoS\footnote{\url{http://www.ioffe.ru/astro/NSG/BSk/}} \citep{2013A&A...560A..48P,2018MNRAS.481.2994P}.
The sound speed is calculated from $c_\mathrm{s}=\sqrt{\mathrm{d}P/\mathrm{d}\rho}$.

The density and sound speed profiles of a $1.4\,M_\odot$ SS and NS are shown in Figure~\ref{fig:cs}.
The SS has a smaller radius ($R_\mathrm{SS}=11.0\,\mathrm{km}$) than the
NS ($R_\mathrm{NS}=12.6\,\mathrm{km}$). Moreover, the SS has a rather uniform density and sound speed
profile with a sharp edge, while the NS's density and sound speed vary significantly from its center
to surface. These differences indicate that the drag force exerted on the inspiraling PBH may be
different in these two cases.

For the parameter $\lambda$, although there are analytical expressions for EoSs when the
adiabatic index is $\Gamma\geqslant1$ \citep{2021PhRvD.103j4009R,2021MNRAS.504.5039A}, no simple
expression is available when $\Gamma<1$ (at the phase transition region of a NS;
\citealt{2013A&A...560A..48P}) and $\Gamma\to\infty$ (at the surface of a SS;
\citealt{2021ChPhC..45e5104X}). So, we use a commonly-used value of $\Gamma=4/3$ in our
calculations, which naturally gives $\lambda=1/\sqrt{2}$ \citep{2020PhRvD.102h3004G,2021MNRAS.504.5039A}.
Such a value is appropriate for most of the density range in NSs and SSs \citep{2013A&A...560A..48P,2021ChPhC..45e5104X}.

\begin{figure}
  \plotone{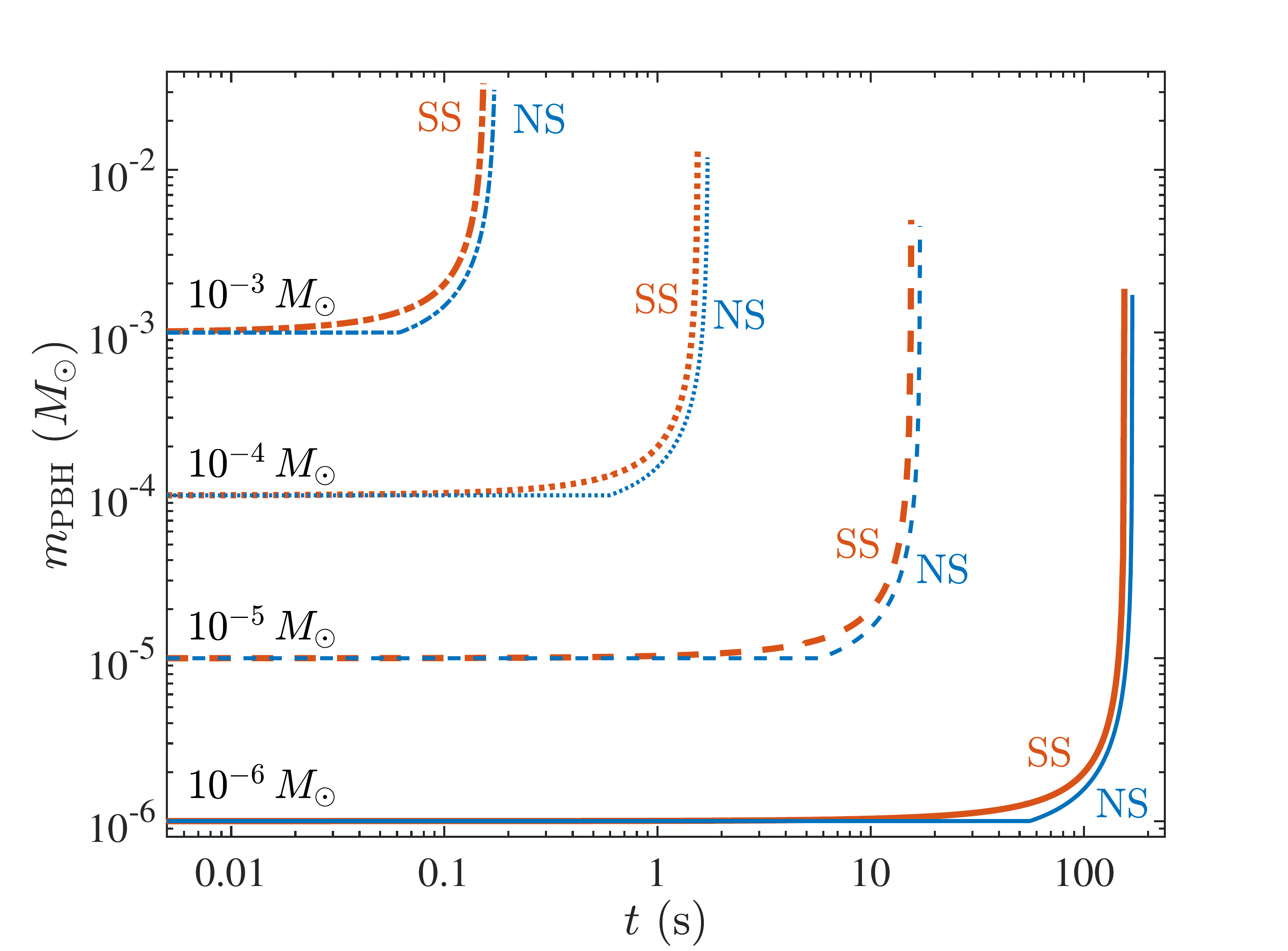}
  \caption{Evolution of the PBH mass. For the dash-dotted, dotted, dashed, and solid curves,
           the initial mass of the PBH is taken as $10^{-3}$, $10^{-4}$, $10^{-5}$,
           and $10^{-6}\,M_\odot$, respectively. The mass of the compact star is 1.4~$M_\odot$,
           which is assumed to be either an SS (thick curves) or an NS (thin curves).\label{fig:m}}
\end{figure}

\subsection{Equation of Motion}

During the inspiral, we assume that the compact star structure inside $r$ stays unchanged
and the whole compact star remains spherically symmetric to its center. Thus the equation
of motion can be written in the relative-motion frame as
\begin{equation}
  \!\!\!\!\!\!\!\!\!\!\!\!\!\!\!\ddot{\boldsymbol{r}} = -\frac{MM_r\boldsymbol{r}}{M_\mathrm{CS}r^3} +
  \frac{M\boldsymbol{F}_\mathrm{DF}}{m_\mathrm{PBH}M_\mathrm{CS}} +
  \frac{\boldsymbol{F}_\mathrm{acc}}{m_\mathrm{PBH}} - \frac{32}{5}
  \frac{(M_r+m_\mathrm{PBH})M_rm_\mathrm{PBH}}{r^4}
  \left[1+\frac{M_r+m_\mathrm{PBH}}{r}\left(-\frac{743}{336}-\frac{11}{4}
  \frac{M_rm_\mathrm{PBH}}{(M_r+m_\mathrm{PBH})^2}\right)\right]\boldsymbol{v},\label{eq:eom}
\end{equation}
where $M=M_\mathrm{CS}+m_\mathrm{PBH}$ is the total mass of the system. The last post-Newtonian
term \citep{2014LRR....17....2B} is introduced to account for the GW energy-loss.

The PBH is initially assumed to be located at the surface of the compact star, with a Keplerian
velocity ($\sqrt{M/R_\mathrm{X}}$, X stands for SS or NS) in a circular orbit. We terminate our
calculation when $m_\mathrm{PBH}/r = 1/12$, which means $r$ reduces to a value comparable to
the innermost stable circular orbit of the PBH and the post-Newtonian method loses its
accuracy \citep{2011mmgr.book.....B}.

\section{Binary Evolution}\label{sec:binary}

\begin{figure}
  \plotone{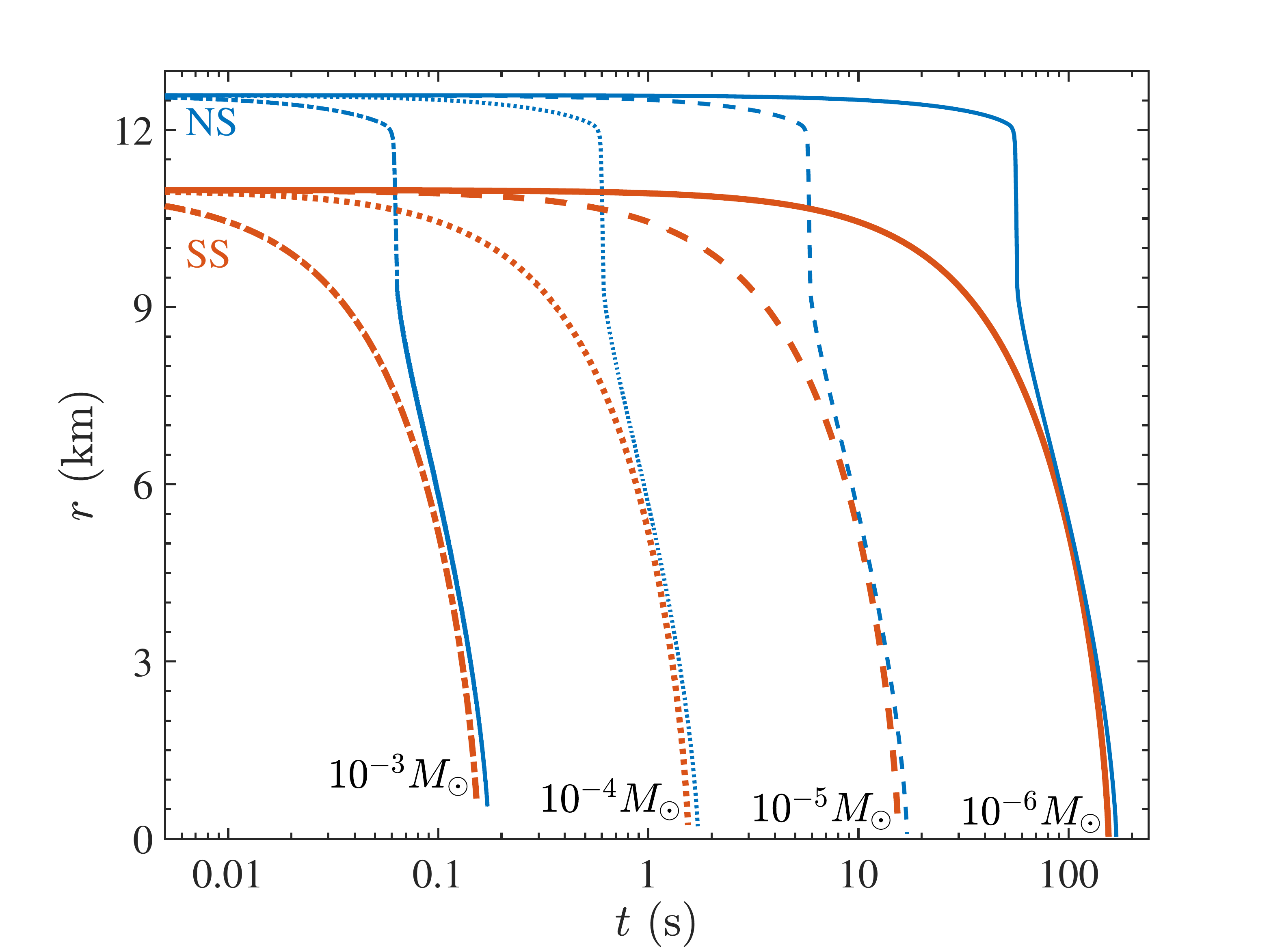}
  \caption{Evolution of the separation between the PBH and the compact star center.
           Line styles are the same as in Figure~\ref{fig:m}.\label{fig:r}}
\end{figure}

We have solved Equation~\ref{eq:eom} numerically. For the initial mass of the PBH, we take
four typical values, i.e., $10^{-3}$, $10^{-4}$, $10^{-5}$, and $10^{-6}\,M_\odot$. As for
the compact star, we assume that it is either an SS or an NS, both with an initial mass
of $M_0=1.4\,M_\odot$. Figure~\ref{fig:m} shows the evolution of the PBH mass. As expected,
a PBH with a larger initial mass accretes faster. Moreover, the mass of the PBH interacting
with an SS increases faster than that of the corresponding NS case. This is due to the
large surface density of the SS. Note that in all our calculations, the condition of
$m_\mathrm{PBH}\ll M_\mathrm{CS}$ is satisfied so that Equation~\ref{eq:accrete} is applicable.

Figure~\ref{fig:r} shows the evolution of the separation between the PBH and the center of the
compact star. We see that the PBHs quickly inspiral inward as its mass increases. At the end of
our calculation, the PBH almost settles at the center of the compact star, which means
the remnant matter of the compact star will soon be completely swallowed by the black
hole \citep{2011mmgr.book.....B}. In the NS cases, the separation decreases sharply in the range
of 9~km~$\lesssim r\lesssim$~12~km. This is because the sound speed and density of the NS
increase quickly in the region. As a result, the Mach number drops to an intermediate value,
leading to an enhancement in the dynamical friction \citep{2007ApJ...665..432K}.

\begin{figure}
  \plotone{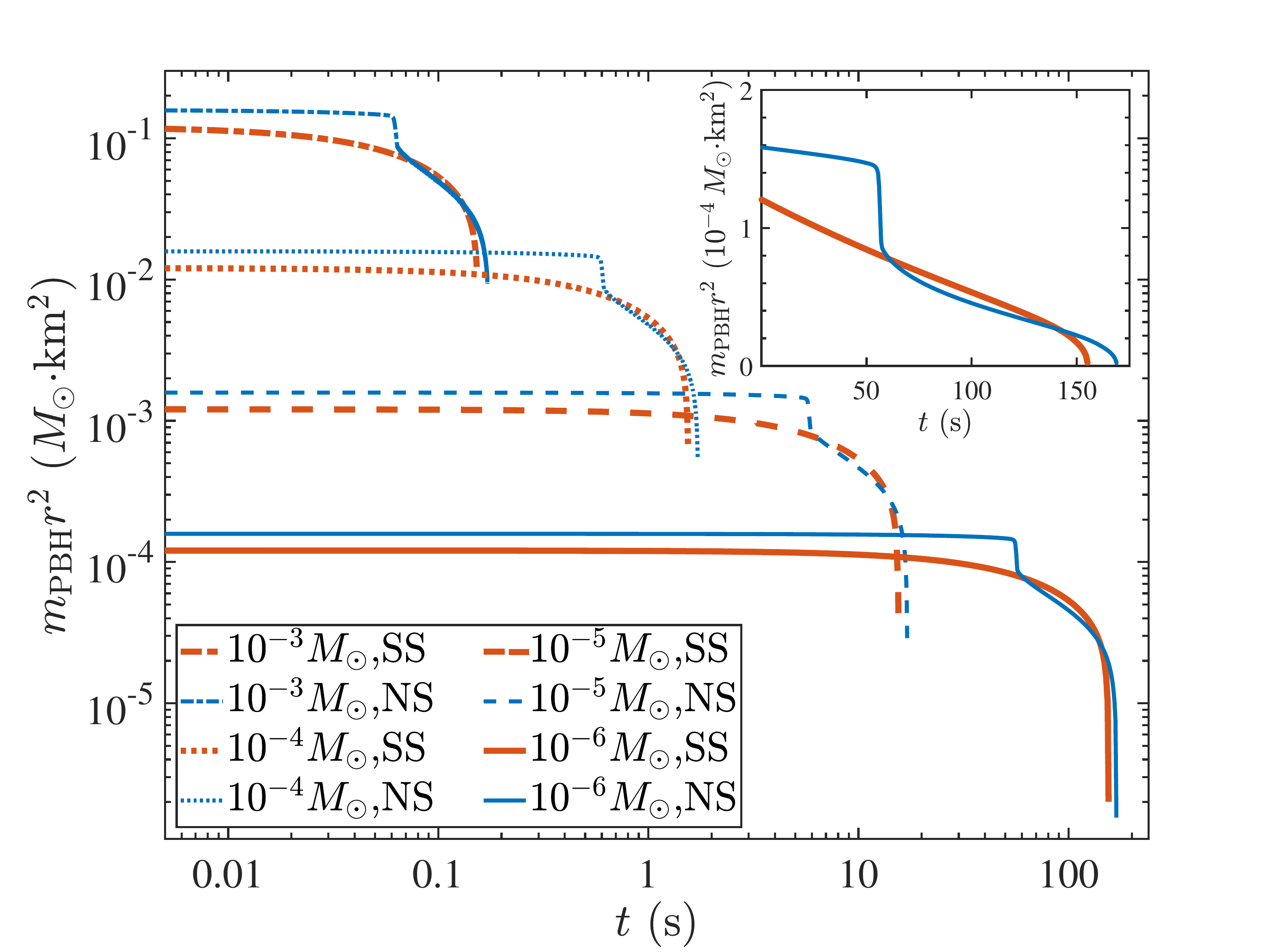}
  \caption{Evolution of $m_\mathrm{PBH}r^2$. The inserted panel at the upper-right
           corner shows the $10^{-6}\,M_\odot$ PBH case on linear scale. Line styles
           are the same as in Figure~\ref{fig:m}.\label{fig:mr2}}
\end{figure}

Figure~\ref{fig:mr2} illustrates the evolution of the quantity $m_\mathrm{PBH}r^2$.
It is interesting to note that \citet{2020PhRvD.102h3004G} suggested that $m_\mathrm{PBH}r^2$ is
an adiabatic invariant during the inspiral. Our Figure~\ref{fig:mr2} shows that
$m_\mathrm{PBH}r^2$ is roughly constant only at the very early stage of the inspiral, while it
generally decreases in the later process. The later decreasing of $m_\mathrm{PBH}r^2$ may be
caused by two reasons. First, the constant-$m_\mathrm{PBH}r^2$ approximation is only valid in the
deeply subsonic regime, but the PBH moves at a speed comparable to or larger than the sound speed.
Second, when the PBH decelerates to a subsonic speed, it becomes very massive and it is also near
the center of the compact star. As a result, the post-Newtonian term dominates the dynamical evolution.

\begin{figure}
  \plotone{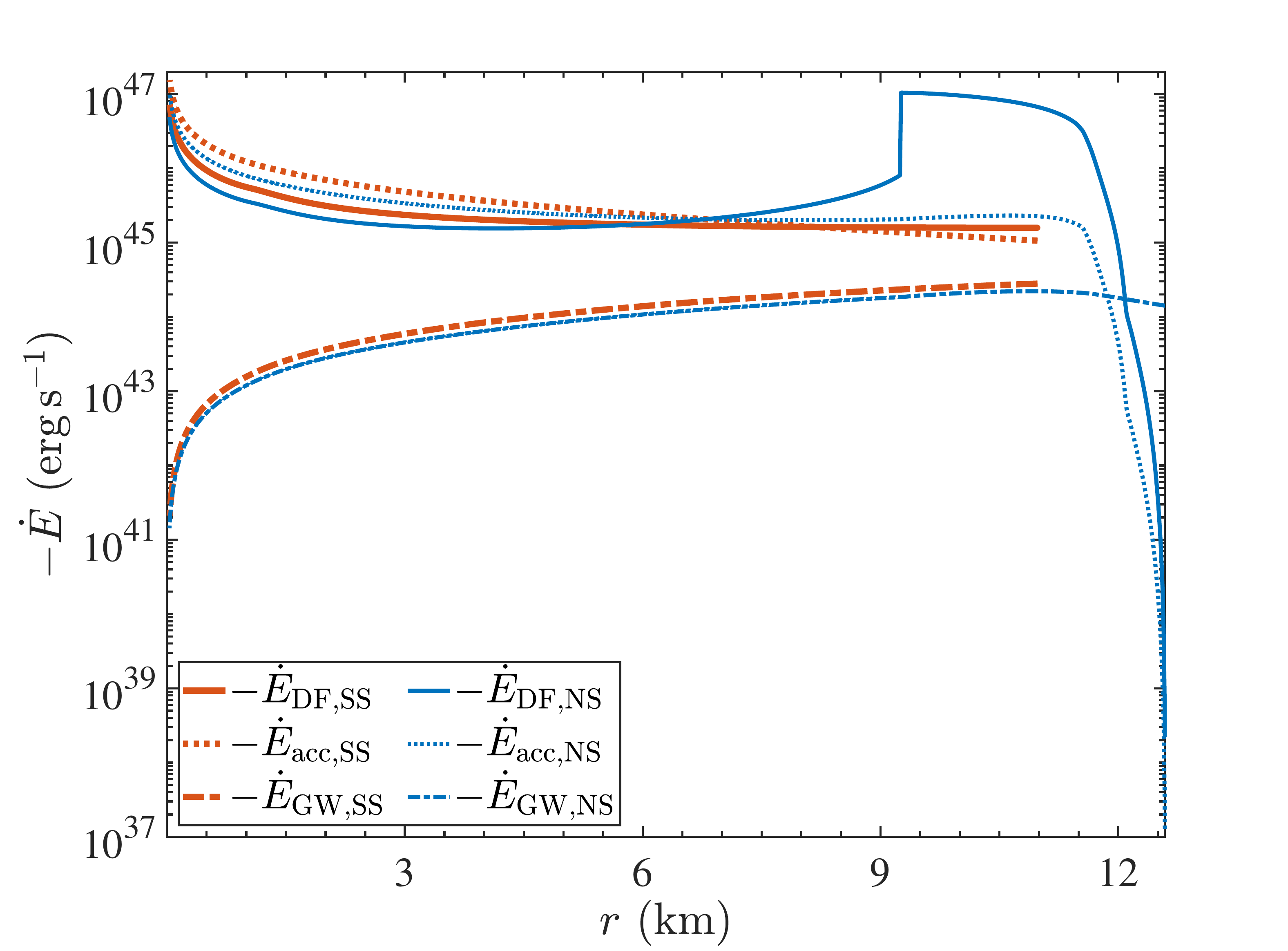}
  \caption{Energy dissipation rates due to dynamical friction (solid curves), accretion (dotted
  curves), and GW emission (dash-dotted curves) at different orbital radius inside an SS (thick curves) or
  NS (thin curves).}\label{fig:edot}
\end{figure}

It would be helpful to compare the individual
contributions of different energy-losing channels. The energy
dissipation rates of the damping forces are
$\dot{E}_\mathrm{DF}=\boldsymbol{F}_\mathrm{DF}\cdot\boldsymbol{v}$
and
$\dot{E}_\mathrm{acc}=\boldsymbol{F}_\mathrm{acc}\cdot\boldsymbol{v}$,
while the quadrupole GW emission has a power of
\citep{maggiore2007gravitational}
\begin{equation}
    \dot{E}_\mathrm{GW}=-\frac{32}{5}\left(\frac{m_\mathrm{PBH}M_\mathrm{CS}}{Mr}v^3\right)^2
    \simeq-\frac{32}{5}\frac{m_\mathrm{PBH}^2v^6}{r^2}.\label{eq:EGW}
\end{equation}
Taking the initial PBH mass as $10^{-6}\,M_\odot$, we have plot
$-\dot{E}$ of the dynamical friction, accretion, and GW emission
against the PBH orbital radius in Figure~\ref{fig:edot}. We can
see that at small radii where the PBH is deeply subsonic, the
accretion dominates, but the dynamical friction is still
non-negligible. At larger radii, the dynamical friction dominates
over accretion, especially in the NS case where the PBH has an
intermediate Mach number. In the NS crust, where the density is
very low, only the GW emission has a significant contribution,
because the dynamical friction and accretion are very ineffective.
Note that Equation~\ref{eq:EGW} underestimates the GW power when
$r$ approaches the innermost stable circular orbit of the PBH
\citep{maggiore2007gravitational,2011mmgr.book.....B}. For all the
three energy-losing channels, the dissipation rate generally
scales as $\dot{E}\propto m_\mathrm{PBH}^2$. Note that when
$\mach\geqslant1$, an additional logarithmic term presents in the
dynamical friction, but the dissipation rate still does not
deviate too much from the $\dot{E}\propto m_\mathrm{PBH}^2$
pattern. As a result, the \emph{relative} dissipation rates of the
three energy-losing channels are almost irrelevant to
$m_\mathrm{PBH}$.

\section{Gravitational Wave Signal}\label{sec:gw}

\begin{figure}
  \plotone{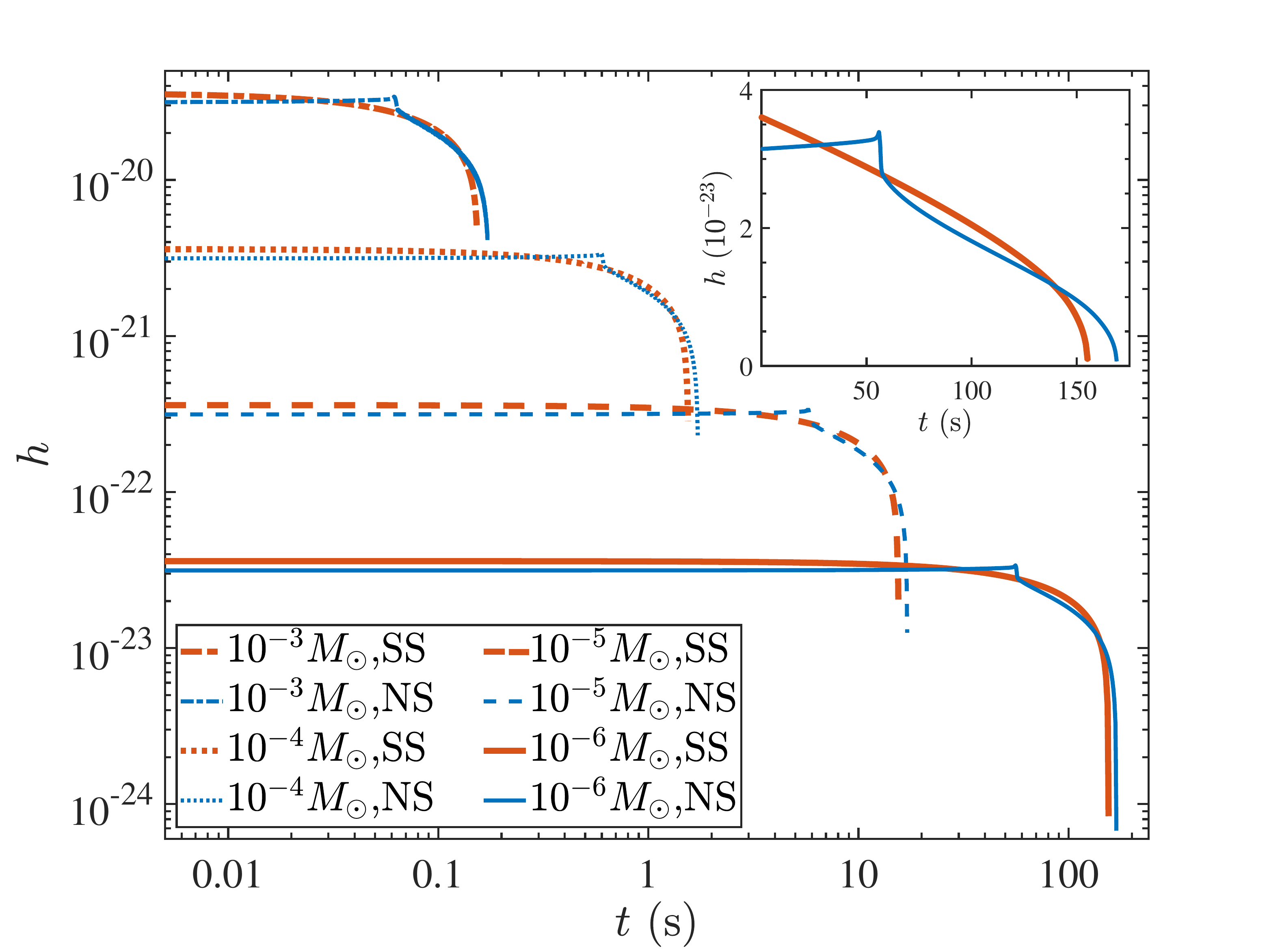}
  \caption{Evolution of the strain amplitude of the gravitational waves. The inserted
           panel at the upper-right corner shows the $10^{-6}\,M_\odot$ PBH case on
           linear scale. Line styles are the same as in Figure~\ref{fig:m}.\label{fig:h}}
\end{figure}

Under the quadrupole approximation, the leading-order GWs of a point mass
moving inside a spherical object are similar to the case of two point
masses \citep{1995A&A...303..789N,2020MNRAS.493.4861G}. As a result, the GW waveforms
of a PBH moving inside a compact star are \citep{2011gwpa.book.....C}
\begin{align}
  h_+&=-\frac{4\mu v^2}{D_\mathrm{L}}\cos2\varphi,\\
  h_\times&=-\frac{4\mu v^2}{D_\mathrm{L}}\sin2\varphi,
\end{align}
where $\mu=m_\mathrm{PBH}M_\mathrm{CS}/M$ is the reduced mass, $D_\mathrm{L}$ is the
luminosity distance, $\varphi$ is the orbital phase. Here the system is assumed to be
face-on. Taking the distance as $D_\mathrm{L} = 1$ kpc, the GW strain amplitude
$h=4\mu v^2/D_\mathrm{L}$ is shown in Figure~\ref{fig:h}. Generally, the strain
amplitude $h$ decays with time. But it is interesting to note that there is
obvious difference between the two curves corresponding to SS and NS.
Thus the GW signal may be used to probe the EoS of dense matter.

\begin{figure*}
  \gridline{\leftfig{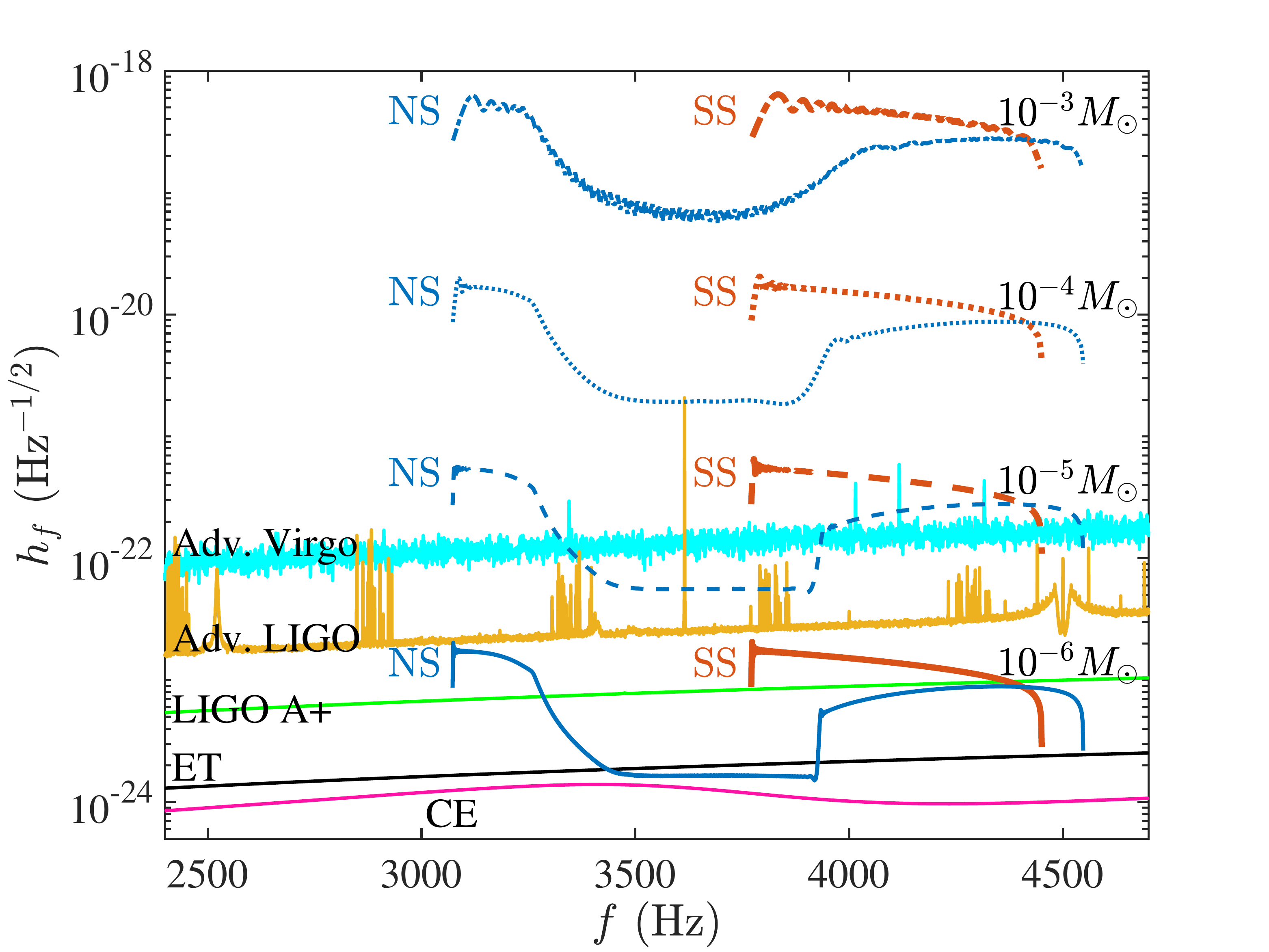}{0.5\textwidth}{(a)}\rightfig{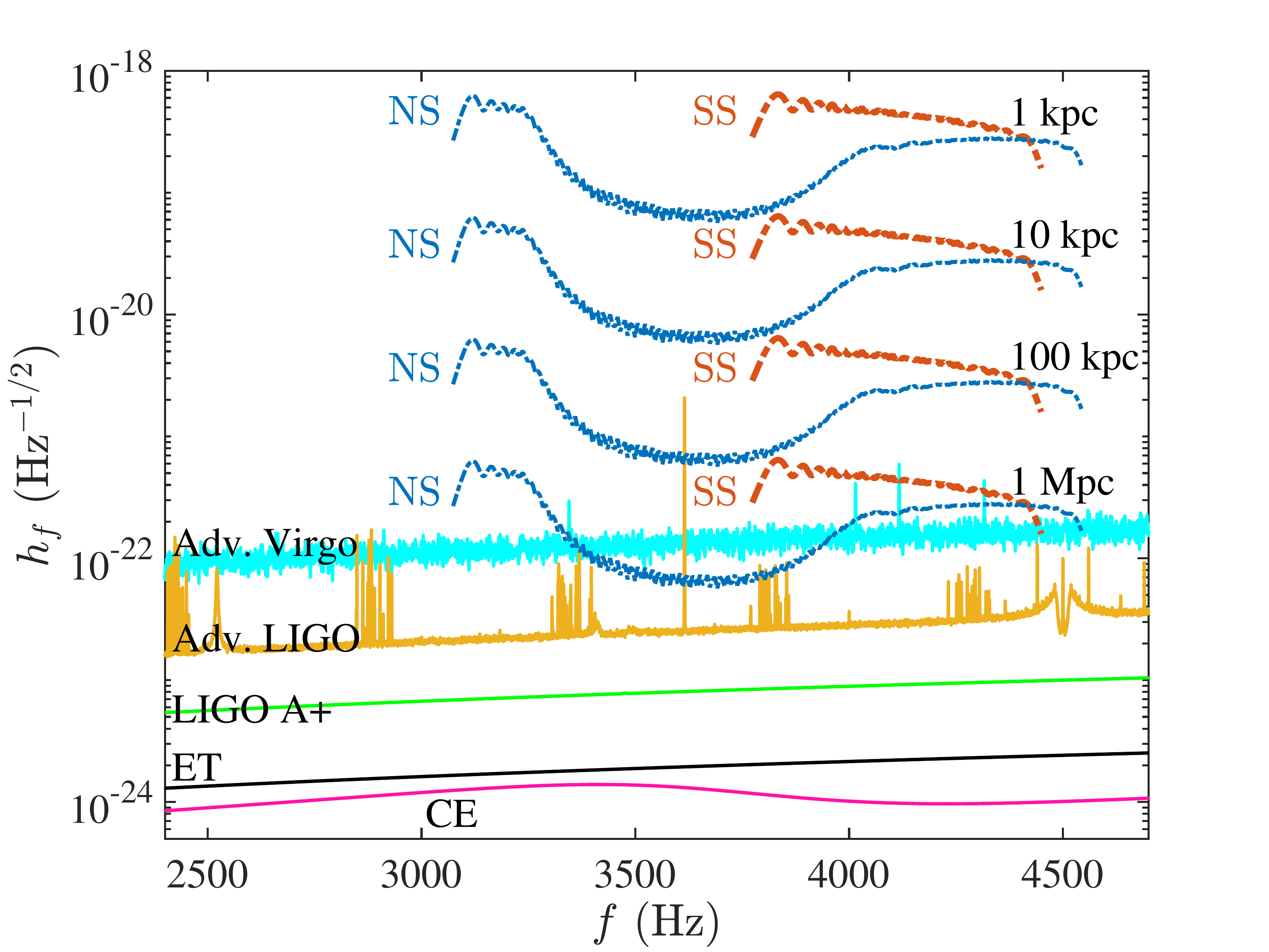}{0.5\textwidth}{(b)}}
  \caption{Strain spectral amplitude of GWs against frequency for a PBH inspiraling inside a compact
           star. In Panel (a), the system is fixed at 1~kpc away, but different initial
           masses are assumed for the PBH and are marked near the corresponding curves.
           In Panel (b), the initial mass of the PBH is fixed as $10^{-3}\,M_\odot$, but the
           systems are put at different $D_\mathrm{L}$ (see the marked values). Sensitivity curves of
           Advanced Virgo, Advanced LIGO, LIGO A+ upgrade, Einstein Telescope (ET), and Cosmic Explorer
           (CE) are plotted for a direct comparison. Other line styles are the same as in
           Figure~\ref{fig:m}.\label{fig:hf}}
\end{figure*}

To decide whether the GW signal can be detected by a particular GW detector or not, it is useful
to calculate the GW strain spectral amplitude \citep{2015CQGra..32a5014M,2020RAA....20..137Z}
\begin{equation}
  h_f=2f^{1/2}\left|\tilde{h}(f)\right|,
\end{equation}
and compare it with the sensitivity of GW detectors. Here $\tilde{h}(f)$ is an average of the
Fourier transforms of $h_+$ and $h_\times$ at frequency $f$. Actually, $h_f$ is the square root
of the GW power spectral density. In Figure~\ref{fig:hf}, we illustrate $h_f$ in a frequency
range of $\sqrt{M_0/R_\mathrm{X}^3}/\uppi\leqslant f\leqslant\sqrt{4\rho_\mathrm{c}/(3\uppi)}$,
where $\rho_\mathrm{c}$ is the central density of the compact star. Beyond this main frequency
range (3 -- 5 kHz) of the GW signals, our calculation of $\tilde{h}(f)$ is significantly
polluted by spectral leakage. Comparing $h_f$ with the sensitivity curves of various GW detectors,
we can see that the GW amplitude from such a PBH system containing a $10^{-5}\,M_\odot$
black hole at 1~kpc away is about one magnitude higher than the sensitivity curve of the
current Advanced LIGO\footnote{\url{https://dcc-lho.ligo.org/LIGO-T2000012/public}}.
It means that the Advanced LIGO and the upcoming LIGO A+ upgrade can safely detect the
inspiraling of a $10^{-5}$ solar mass primordial black hole at a distance of 10 kpc.
For a Jovian-mass PBH, due to the much stronger GW emission, the detection horizon can even
be pushed to megaparsecs. The next generation detectors such as Einstein
Telescope\footnote{\url{http://www.et-gw.eu/index.php/etsensitivities}} and Cosmic
Explorer\footnote{\url{https://cosmicexplorer.org/researchers.html}}, with sensitivities almost two
magnitudes better, can even detect the GWs from an inspiraling $10^{-5}\,M_\odot$ PBH at $\sim 1$ Mpc,
and detect Jovian-mass cases at several hundred megaparsecs.

Figure~\ref{fig:hf} clearly shows that although GWs in the NS case and the SS case have similar
amplitudes, the shape of the detailed $h_f$ curves actually is quite different for these two EoSs.
For example, $h_f$ in the SS cases generally decreases homogeneously with the increase of the frequency,
while $h_f$ in the NS cases shows a concavity at intermediate frequencies. As a result, GW detectors
working in kilohertzes can hopefully help us probe the EoS of dense matter through signals from
PBHs inspiraling inside compact stars.

\section{Effects of Eccentricity}\label{sec:e}

The spatial speeds of PBHs may distribute in a wide range.
High-speed PBHs could not be captured by a compact star. Only
those PBHs with a relative speed small enough would be
successfully captured by a compact star. For simplicity, the PBH
is assumed to be initially in a circular orbit in our
calculations. However, a PBH could be accelerated to a speed
significantly larger than the circular Keplerian speed when it is
gravitationally captured by the compact star, thus an orbit with a
large eccentricity is essentially possible
\citep{2020PhRvD.102h3004G}. At such an early stage of the
capture, the system does not produce inspiraling GW signals as
described in Section~\ref{sec:gw}, but generates intermittent GW
emissions at each passage of the perihelion as pointed out by
\citet{2020PhRvD.102h3004G}.

If the perihelion is located outside the compact star, the
eccentricity will gradually decrease due to GW emission
\citep{1964PhRv..136.1224P,2011mmgr.book.....B,2011gwpa.book.....C,2014LRR....17....2B}.
Moreover, the tidal dissipation can also circularize the orbit
\citep{2014ARA&A..52..171O}. In most cases, the circularization
will be sufficient and the PBH enters the compact star in a
quasi-circular orbit.

On the other hand, if the PBH enters the compact star with
a substantial eccentricity, or even its perihelion is initially
located inside the compact star, it will lose part of its kinetic
energy due to the interaction with the compact star matter near
the perihelion, and will be finally trapped inside the compact
star after several passages \citep{2018ApJ...868...17A}. When the
PBH moves completely inside the compact star, the accretion
process can efficiently circularize the orbit
\citep{2013ApJ...774...48M}, especially for the extreme mass ratio
cases considered in this work
($m_\mathrm{PBH}/M_\mathrm{CS}\sim10^{-6}-10^{-3}$). Therefore,
our previous calculations of the dynamics and GW emissions are
still valid for the later stage of a PBH captured by a compact
star. For the earlier stage when the trapped PBH still has a
non-negligible eccentricity, Equation~\ref{eq:eom} can be used to
calculate the motion of the PBH, but the dynamical friction term
and the resulting GW waveform will be very different.

\section{Event Rate}\label{sec:rate}

If PBHs account for a fraction $\mathcal{X}$ of the dark
matter, the detectable GW event rate of compact stars capturing
PBHs can be calculated by
\citep{2013PhRvD..87l3524C,2017PhRvL.119f1101F}
\begin{equation}
    F=\sqrt{6\uppi}\mathcal{X}N
    \frac{\rho_\mathrm{DM}R_\mathrm{S}R_\mathrm{X}}
    {m_\mathrm{PBH}\bar{v}\left(1-R_\mathrm{S}/R_\mathrm{X}\right)}
    \left\{1-\exp\left[-3E_\mathrm{loss}/\left(m_\mathrm{PBH}\bar{v}^2\right)\right]\right\},
    \label{eq:rate}
\end{equation}
where $\rho_\mathrm{DM}$ is the density of dark matter, $N$ is the
number of compact stars within GW detectors' horizon,
$R_\mathrm{S}$ is the Schwarzschild radius of compact stars, and
$E_\mathrm{loss}\propto m_\mathrm{PBH}^2$ is the energy loss
during the PBH-compact star interaction. Here we have assumed a
Maxwellian distribution for the PBH velocities, with a dispersion
of $\bar{v}$. For the PBH mass, we take a typical value of
$m_\mathrm{PBH} = 10^{-5}\,M_\odot$ in our calculations below.

There are 281 pulsars within the 1 kpc range of the
Sun\footnote{\url{https://www.atnf.csiro.au/people/pulsar/psrcat/}}
\citep{2005AJ....129.1993M}. Considering that pulsars have an
average lifetime of $\sim10^7$ years (compared to the lifetime of
$\sim10^{10}$ years of the Milky Way) and beaming factor of
$\sim3.5$ \citep{2012puas.book.....L}, the true number of compact
stars within 1 kpc of the Sun can be estimated as $\sim10^6$.
Since the GW amplitude of an inspiraling $10^{-5}\,M_\odot$ PBH at
1 kpc away is about one magnitude higher than the sensitivity curve of
the Advanced LIGO (see Figure~\ref{fig:hf} (a)), the Advanced LIGO's
horizon is therefore $\sim10$ kpc, which nearly includes the
whole Milky Way. As a result, there are $N\sim10^9$ compact stars
in the detection horizon. This is a logical number, which actually
coincides with the total number of compact stars in the Milky Way.
For PBHs with $m_\mathrm{PBH}\sim10^{-5}\,M_\odot$, one has
$3E_\mathrm{loss}\gg m_\mathrm{PBH}\bar{v}^2$
\citep{2013PhRvD..87l3524C}, thus the exponential term in
Equation~\ref{eq:rate} can be neglected. Adopting a radius of
$R_\mathrm{X}\sim12\,\mathrm{km}$ for a typical 1.4\,$M_\odot$
compact star (see Section~\ref{subsec:CS}) and taking the PBH
velocity dispersion as $\bar{v}\sim105\,\mathrm{km\,s}^{-1}$
\citep{2017PhRvL.119f1101F}, the detectable event rate within a
10~kpc horizon (by the Advanced LIGO) is
\begin{equation}
    F\sim3.2\times10^{-12}\,\mathrm{year}^{-1}\frac{\mathcal{X}}{0.01}\cdot
    \frac{N}{10^9}\cdot\frac{10^{-5}\,M_\odot}{m_\mathrm{PBH}}\cdot
    \frac{\rho_\mathrm{DM}}{0.4\,\mathrm{GeV\,cm}^{-3}}\cdot
    \frac{105\,\mathrm{km\,s}^{-1}}{\bar{v}},
    \label{eq:1kpcrate}
\end{equation}
where $\rho_\mathrm{DM}=0.4\,\mathrm{GeV\,cm}^{-3}$ is the local
dark matter density \citep{2021RPPh...84j4901D}. This rate seems
to be very small. However, note that in Equation~\ref{eq:1kpcrate}
we have assumed that both the PBHs and the compact stars
distribute homogeneously in a spherical volume with a radius
of 10 kpc.

In a realistic case, both the dark matter and the compact
stars concentrate at the Galactic center, and the high densities
will markedly increase the event rate. The Galactic center is
$\sim 8$ kpc away from us. In fact, the dark matter density at the
Galactic center might be as high as
$\rho_\mathrm{DM}=10^6\,\mathrm{GeV\,cm}^{-3}$
\citep{2005MPLA...20.1021B}, and a population of up to $N \sim 10^8$
compact stars may reside there
\citep{1997ApJ...475..557C}. In such a dense environment,
multi-body interactions can further increase the capture rate by a factor
of $\sim3.5$ \citep{2012PhRvL.109f1301B}. Moreover, note that
Equation~\ref{eq:rate} only takes the dynamical friction into
account \citep{2013PhRvD..87l3524C}, while the energy loss through
GW emission can additionally enhance the capture rate by up to a factor
of 10 for PBHs with $m_\mathrm{PBH}\gtrsim10^{-6}\,M_\odot$
\citep{2020PhRvD.102h3004G}. Considering all the above factors and
assuming PBHs account for a portion of 0.01 -- 0.1 of the dark matter at the
Galactic center \citep{2019PhRvD..99h3503N}, the event rate solely from the
Galactic center can be estimated as $F\sim10^{-5}$ -- $10^{-4}\,\mathrm{year}^{-1}$.
This rate is rather low, but still nonzero. It is also consistent with
the nondetection of such events by the LIGO-Virgo Collaboration up until now.

Future GW detectors like the Einstein Telescope and the Cosmic Explorer
have sensitivities nearly two magnitudes better than the advanced LIGO
(see Figure~\ref{fig:hf}), thus may have a horizon of $\sim 1$ Mpc.
About 10 -- 100 galaxies may reside in this range, leading to a
total detectable event rate of $F\sim10^{-4}$ -- $10^{-2}\,\mathrm{year}^{-1}$.
Note that there are still other factors that can further enhance the
event rate. For example, globular clusters are expected to have a
relatively high dark matter density as well as a low velocity
dispersion. Thus they are also interesting places for PBH captures
\citep{2022PhRvL.128b1101D}. Moreover, the tidal effect \citep{2020PhRvD.102h3004G}
and possible underestimation of $\rho_\mathrm{DM}$ and $N$ at the Galactic
center can also increase the event rate, which are however beyond the
scope of this study. Anyway, the upcoming next-generation
GW detectors may hopefully succeed in detecting such events.

\section{Summary and Discussion}\label{sec:conclude}

The process of a planetary-mass PBH inspiraling inside a compact star is investigated in detail.
Such a process is of great interest for constraining PBH's composition fraction of dark matter.
The effects of dynamical friction, accretion, and GW emission on the motion of the PBH are
taken into account. Two kinds of compact stars are considered, i.e., SSs and normal NSs.
It is found that the resulting GW signals show significant difference between the two cases.
Encouragingly, the current Advanced LIGO detector can detect a $10^{-5}\,M_\odot$
PBH case at 10 kpc away from us, and detect a Jovian-mass PBH case up to
megaparsecs. The planned Einstein Telescope and Cosmic Explorer
can detect $10^{-5}\,M_\odot$ PBHs at $\sim 1$ Mpc, and detect Jovian-mass PBHs
at several hundred megaparsecs. Note that the GW frequencies are near the high end of
the sensitivity curve for most ground-based interferometers. Future ad hoc detectors working
at kilohertz frequencies like Neutron Star Extreme Matter Observatory \citep{2020PASA...37...47A}
will be powerful tools for probing the EoSs of dense matter.

Our dynamical equation (Equation~\ref{eq:eom}) is invalid at the final stage when the whole compact
star is to be swallowed by the PBH to form a stellar mass black hole. In our study, we did not
calculate the GWs emitted at this final stage. However, since the PBH has moved to the central
region of the compact star at the end of our calculation, the accretion by the black hole should
be highly spherical and the GWs are expected to be much weaker according to Birkhoff's theorem.
In our modeling, we have omitted the spinning of the compact star for simplicity. But note that
the rotation of the compact star would not affect the results significantly, because even a
millisecond pulsar rotates much slower than its Keplerian velocity \citep{2020PhRvD.102h3004G}.
Another approximation is that a constant $\lambda$ is adopted during our calculations. To overcome
this weakness, the GW signals should be calculated through full general relativistic hydrodynamic
simulations, which is beyond the scope of the current study.

Before the PBH contacts the NS/SS surface and begins its journey inside the compact star, it orbits
around its host as a very close-in object, which will also emit strong GWs. Such GWs may be recognized
as originating from a ``planet''-compact star system. \citet{2015ApJ...804...21G} and
\citet{2020ApJ...890...41K} have suggested that GWs from such ``planet''-compact star systems could be
efficiently used to identify strange quark planets, because a normal matter planet will be tidally disrupted
by the compact star when it is still further away so that no GWs are available
\citep{2015ApJ...804...21G,2020ApJ...890...41K}. Here, we would like to further remind that the possibility
that the ``planet'' is actually a PBH should be further repelled before finally identifying it as a strange
quark planet. The GWs emitted by the PBH inspiraling inside its host compact star, as calculated by us in
this study, can help us with this task. The merging of a strange quark planet with an SS will only produce
some kind of ringing-down in GWs, while a PBH will tunnel through its host and produce complicated GW
patterns as demonstrated in this study.

PBHs colliding with compact stars are also hypothetically associated with many other interesting
phenomena, including the formation of solar mass black holes \citep{2021PhRvL.126g1101T} and
fast radio bursts \citep{2018ApJ...868...17A,PhysRevD.104.123033}. Future observations of the GWs
will help constrain these hypotheses.

\begin{acknowledgments}

We thank the anonymous referees for valuable suggestions that led
to an overall improvement of this study. We acknowledge the ATNF
Pulsar Catalogue for useful data. This work is supported by the
National Natural Science Foundation of China (Grant Nos. 12041306,
11873030, U1938201), by the National Key R\&D Program of China
(2021YFA0718500), by National SKA Program of China No.
2020SKA0120300, and by the science research grants from the China
Manned Space Project with NO. CMS-CSST-2021-B11.

\end{acknowledgments}

\bibliographystyle{aasjournal}

\end{document}